\documentstyle[12pt,aaspp4]{article}

\lefthead{Bekki Kenji}
\righthead{Merging between black holes  and stellar systems}

\begin{document}
\title{Merging between a central massive black hole and
a compact stellar system:
A clue to the origin of  M31's nucleus}

\author{Kenji Bekki} 
\affil{Division of Theoretical Astrophysics,
National Astronomical Observatory, Mitaka, Tokyo, 181-8588, Japan} 

\begin{abstract}

The central bulge of M31 is observed to have two distinct brightness
peaks with the separation of $\sim$ 2 pc.  
Tremaine (1995) recently proposed a new idea that
the M31's nucleus is  actually a single thick eccentric disk
surrounding the central super-massive black hole.
In order to explore the origin of the proposed eccentric disk,
we numerically investigate 
the dynamical evolution of a merger between a central massive black
hole with the mass of $\sim$ $10^7$ $M_{\odot}$ and a compact stellar system
with the mass of $\sim$ $10^6$ $M_{\odot}$ and the size of
a few  pc in the central 10 pc of a galactic bulge.
We found that the  stellar system
is  destroyed by
strong tidal field of the massive black hole
and consequently
forms a rotating nuclear thick stellar disk.  
The orbit of each stellar component in the developed disk
is rather eccentric with the mean eccentricity of $\sim$ 0.5. 
These results imply that the M31's nuclear eccentric disk 
proposed by Tremaine (1995)
can be  formed by merging between a  central massive black hole
and a  compact stellar system.
We furthermore discuss when and how a compact stellar
system is transferred into the nuclear region around  a massive
black hole. 

\end{abstract}


\section{Introduction}
Since high-resolution photographs by the Stratoscope 
II balloon-borne telescope first resolved the nucleus of M31
(Light,  Danielson,  \& Schwarzshild 1974),
the origin of the peculiar nature of the M31's nucleus
has been extensively investigated  both by  observationally 
and theoretically.
In particular, the asymmetry of the central brightness peak
of M31 was observationally investigated 
in various bands and the origin of the asymmetry
was extensively discussed (e.g.,
Light et al. 1974, Nieto et al. 1986; Mould et al. 1989;
Bacon et al. 1994). 
Recent observational studies by the Hubble Space Telescope ($HST$) 
have revealed that the M31's central bulge has two distinct
brightness peaks with the separation of about 1.7 pc (Lauer et al. 1993;
King, Stanford, \& Crane 1995).
The component with the lower surface brightness (represented by $P2$)
coincides with the bulge photocenter 
whereas the higher surface-brightness (and off-center
nuclear)  component ($P1$) coincides
with the nuclear core revealed by the Stratoscope
(Lauer et al. 1993; King et al.  1995).
Furthermore a growing number of spectroscopic studies have
been accumulated which can reveal the detailed kinematics
of the M31's nuclear region 
(e.g., Dressler \& Richstone 1988; Kormendy 1988; Bacon et al. 1994;
van der Marel et al. 1994).
One of the most remarkable  result of these is that 
although the nuclear rotation curve is symmetric with the peak nearly
coincident
with $P2$, the profile of the velocity dispersion is not symmetric 
(e.g., Bacon et al. 1994).

Although a few ideas 
(e.g., partial dust obscuration and a separated
$P1$ stellar system) are suggested  
by Lauer et al. (1993) and  Bacon et al. (1994)
for plausible explanations
of the asymmetries observed in M31's nucleus, 
these are suggested to be implausible from simple dynamical
arguments  (Tremaine 1995).
Recently, Tremaine (1995) has proposed that
the nucleus of M31 is actually a thick eccentric disk composed
of stars traveling on nearly Keplerian orbits
around a massive black hole (or dark compact object).
Tremaine (1995) furthermore suggested that this eccentric disk
model can clearly explain the rotation curve and asymmetric dispersion
profile revealed by recent grand-based spectroscopic observations.
Kormendy \& Bender (1999) 
confirmed that the Tremaine's model is very consistent with
M31's structural and kinematical properties reveled by $HST$ photometric
studies and grand-based spectroscopic ones. 
Although several ideas as to why the M31's nuclear disk should be
eccentric are discussed (Tremaine 1995), 
it remains highly uncertain how such a stellar  disk can be
formed in the nuclear region of the M31's bulge.

The purpose of this letter 
is to investigate numerically  merging between 
a central massive black
hole with the mass of 
$\sim$ $10^7$ $M_{\odot}$ and a compact stellar system
with the mass of $\sim$ $10^6$ $M_{\odot}$ and the size of
a few  pc.
We  demonstrate  that tidal disruption
of the stellar system by the massive black hole  
in the nuclear region of M31
is important for the formation
of the M31's nuclear eccentric disk proposed by Tremaine (1995). 
We also show eccentricity distribution of stars in
the developed disk and structural and kinematical properties
of the disk.
Not all of the fundamental structural and 
kinematical properties of the M31's nuclei can be clearly
explained by the present model owing to some limitations
of the model.  
However, we consider that
the present study
can provide a new clue to the origin
of the proposed nuclear eccentric disk,
because there are no extensive numerical studies
addressing the origin of the suggested nuclear thick  disk
of M31.

\section{Model}

We consider a purely dissipationless
merger between a central massive black hole
(hereafter referred to as MBH) with the mass of $M_{\rm BH}$
and a compact stellar system (simply as a cluster) with
the mass of $M_{\rm st}$ in the central region
of a galactic bulge.
Both the MBH and the cluster are assumed to feel the fixed 
external gravitational field of the bulge component.
The MBH is assumed to be initially located at the center
of the bulge and the initial separation between the MBH
and the cluster is set to be $R_{\rm ini}$.
From now on, all the mass and length are measured in units of
$M_{\rm BH}$ and  $R_{\rm ini}$, respectively, unless specified. 
Velocity and time are
 measured in units of $v$ = $ (GM_{\rm BH}/R_{\rm ini})^{1/2}$ and
$t_{\rm dyn}$ = $(R_{\rm ini}^{3}/GM_{\rm BH})^{1/2}$, respectively,
where $G$ is the gravitational constant and assumed to be 1.0
 in the present study.
If we adopt $M_{\rm BH}$ =  $10^{7}$ ($10^{8}$) $\rm M_{\odot}$ and
$R_{\rm ini}$ = 10  pc as a fiducial value, then $v$ = 6.5 $\times$
10 (2.1 $\times$ $10^2$) km/s  and  $t_{\rm dyn}$ = 1.49 $\times$ $10^{5}$ 
(4.72 $\times$ $10^4$)  yr,
respectively.
For the radial density profile of the bulge, 
we adopt the universal profile proposed by Navarro, Frenk,
\& White (1996). 
We assume that the scale length (or the characteristic
radius, $r_{\rm s}$) is equal to
 10$R_{\rm ini}$ and determine the central density
so that total mass of the bulge within 200$R_{\rm ini}$ ($\sim$ 2 kpc)
is 200$M_{\rm BH}$. 
The adopted ratio of bulge mass to MBH one is well within
a reasonable value derived by Faber et al. (1997). 
The total mass of the bulge within $R_{\rm ini}$ is hereafter represented by
$M_{\rm gal}$.
We use the so-called Plummer model with the scale length of 0.04$R_{\rm ini}$ 
for the initial density profile of the cluster.
For the fiducial model, the mass ratio  $M_{\rm st}$/$M_{\rm BH}$
and $M_{\rm gal}$/$M_{\rm BH}$ are set to be 0.1 and 0.3, respectively.

The initial orbital plane of the merger
is  assumed to be exactly the same as the $x$-$y$ plane.
Initial $x$ and $y$ position ($x$, $y$) is set to be
(0,0) for the MBH and (1,0) for the cluster in all models.
Initial $x$ and $y$ velocity ($V_{\rm x}$, $V_{\rm y}$)
is set to be (0,0) for the MBH and 
($-0.5V_{\rm cir}$, 0.75$V_{\rm cir}$) for the cluster 
in the fiducial model, where $V_{\rm cir}$ is the circular velocity
(1.14 in our units) at the radius of $R_{\rm ini}$.
The  number of particles 
for  the cluster is 10000 and  
the parameter of gravitational softening is set to be fixed at
4.7 $\times$ $10^{-3}$ in our units 
for all the simulations. 
All the calculations related to 
the above dynamical evolution
have been carried out on the GRAPE board
 (\cite{sug90})
at Astronomical Institute of Tohoku University.
Using the above model, we mainly describe  structural and kinematical
properties of the merger remnant in the fiducial model.
We furthermore investigate the distribution 
of orbital eccentricity ($e$) of the merger remnant
in order to confirm whether the remnant is 
an  eccentric disk.
Here $e$ for each stellar particle is defined as:
\begin{equation}
  e= \frac{r_{apo}-r_{peri}}{r_{apo}+r_{peri}} \;
\end{equation}
where $r_{apo}$ and $r_{peri}$ are apo-galactic and peri-galactic distances
from the center of the bulge, respectively.
Moreover we summarize briefly the dependence of mass distribution
of merger remnant on the initial merger parameters such as $M_{\rm BH}$,
$V_{\rm x}$, and $V_{\rm y}$.
More details on the parameter dependences will be described in our
future papers.

\placefigure{fig-1}
\placefigure{fig-2}
\placefigure{fig-3}

\section{Result}

Figure 1 shows the time evolution of mass distribution 
of a stellar cluster in the present merger model.
As the cluster becomes close to the MBH,
some fraction of stars are tidally stripped away from
the cluster and consequently form  inner very low-density  
eccentric ring-like structure ($T$ = 2.0).
When the cluster passes by the pericenter of the merger orbit,
it suffers severely from strong tidal gravitational
field of the MBH and is consequently distorted greatly ($T$ = 3.2).
As a result of this, the cluster spherical shape
is transformed  into a thick flattened disk with the morphology
looking like a crescent ($T$ = 3.2).
During this morphological transformation,
the initial orbital angular momentum of the merging cluster
is efficiently changed into the intrinsic angular momentum
of the forming disk around the MBH.
Finally a rotating thick stellar disk with a small central
hole (i.e., a central very low-density region)
is formed around the MBH ($T$ = 4.0).
The vertical scale height of the developed thick disk
depends on the initial half-mass radius of the cluster.

As is shown in the upper left panel of Figure 2, the orbits  of stars 
are more likely to be rather elongated and the mean
orbital eccentricity is estimated to be about 0.55.
Most of stars located at $R$ $<$ $R_{\rm ini}$ (= 1.0),
where $R$ is the distance from the center of the bulge,
show
$e$ $>$ 0.5 and accordingly form the  peak
around 0.8 in the distribution. 
Stars at $R$ $>$ $R_{\rm ini}$, on the other hand,
show moderately small eccentricity ($e$ $<$ 0.3). 
These results clearly demonstrate that the inner thick disk
formed by merging is composed mainly of stars with eccentric orbits. 
As is shown in Figure 2 for $T$ = 4.0,
the developed disk shows strong asymmetry in
structural and kinematical properties.
The radial density profile shows double peaks owing to the hole
formed during tidal interaction between the MBH and the cluster. 
The rational velocity rapidly increases along $x$ axis for 0 $<$ $R$ $<$
0.2,  rapidly decreases for 0.2 $<$ $R$ $<$ 0.6,
and again increases gradually for 0.6 $<$ $R$.
The peak of the velocity dispersion does not coincide with
the center of the bulge (i.e., ($x$,$y$) = (0,0)):
The peak is close to
the location of the MBH (i.e., ($x$,$y$) = (0.9,0.14))
rather than to the bulge center at $T$ = 4.0.
The derived structural and kinematical properties
of the nuclear disk are qualitatively similar to those
observed by Kormendy \& Bender (1999) for M31.

Figure 3 briefly
summarizes the parameter dependences of final mass distribution
of mergers.
Models described in Figure 3 all fail to form a nuclear thick disk
after merging and tidal interaction
and thus show the following four physical conditions
required for the formation of the disk around MBH.
Firstly, $M_{\rm BH}$ should be much ($\sim$ 10 times) larger than 
$M_{\rm st}$ so that MBH can tidally disrupt a cluster during merging.
Secondly, the initial orbit of a cluster
should be rather elongated (i.e., small pericenter distance)
so that a cluster  can be  well within the tidal radius of MBH
during tidal interaction.
Thirdly, 
merging rather than simple
tidal interaction between
a MBH and a cluster is necessary (We here note that even for the 
tidal interaction,
an eccentric ring-like object can be formed).
Fourthly, initial impact parameter (or pericenter distance) 
should not be too small so that the cluster can not be completely
destroyed and randomly dispersed.
These required  conditions imply that even if
merging or tidal interaction between a MBH and a cluster
are frequently occurred in the central region of a bulge,
the bulge can  not necessarily contain a nuclear thick stellar
disk.

\section{Discussion}

 Although the present study showed  one possible mechanism for
the formation of a thick stellar disk around the MBH 
of M31,
there is one  important remaining question:  
What is the progenitor of a compact stellar system that
can finally evolve into the M31's nuclear disk? 
We here suggest the following three promising candidates. 
The first is an old and relatively metal-poor globular cluster system. 
Tremaine, Ostriker, \& Spitzer (1975)
demonstrated  that globular clusters passing near the center
of M31 spiral in to the center of M31 owing
to dynamical friction and consequently are  tidally
disrupted there to form a  distinct high-density nucleus.
This demonstration  leads us to propose  that
a globular cluster spiraling in to the surrounding of 
MBH is a likely candidate
for the progenitor stellar system.
The second is a massive clump that is composed of gas and stars
and is developed in the M31's disk.
Shlosman \& Noguchi (1993) found that if
the gas mass fraction of a globally unstable disk is larger
than 0.1,  massive gas clumps (with the masses of $\sim$ $10^7$ $M_{\odot}$)
formed from local gravitational instability can be transferred into
the central inner kpc owing to dynamical friction.
Noguchi (1998) furthermore demonstrated that these massive clumps
are more likely to be formed in the early disk formation phase
when disk galaxies have a larger amount of gas.
These numerical studies imply that if star formation very efficiently proceeds
in the gas clumps, compact stellar system can reach the surrounding of
the M31's MBH. 

The third candidate is a young massive star cluster 
that is similar to those observed in infrared luminous major mergers
such as Arp 220. 
Shaya et al. (1994) revealed  a number of bright and possibly young
star clusters in the core of Arp 220 and suggested that these
clusters can be very quickly transferred to the inner tens pc
within an order
of $10^8$ years owing to dynamical friction.
Based on metallicity distribution and global
structure of M31's stellar halo, 
Freeman (1999) suggested that M31's bulge was formed by the past merging
events.
Accordingly, it could not be unreasonable to say
that young massive clusters newly created 
in the epoch of M31's bulge formation by major merging 
(at relatively high redshift)
can be transferred  to the inner pc region around the MBH.
Since the nature of stellar population
can be greatly different between the above three candidates
(very old and metal-poor for the first, young and metal-rich for the second,
and relatively old and metal-rich for the third),
the most plausible and realistic candidate of the progenitor 
can be determined
if age and metallicity distribution of stellar components
in the M31's nuclear disk is investigated in detail. 
Kormendy \& Bender (1999) found that the stellar population
of the $P1$ nucleus is more similar to that of $P2$ than it is
to the bulge or to a globular cluster.
This result strongly suggests that it is unlikely that
the M31's eccentric disk consists  of accreted old stars.
Lauer et al. (1998) furthermore revealed that
the $P2$ nucleus shows the spectral energy distribution
consistent with late B-early A stars  
and thus can  have relatively young stellar populations. 
Although
these observational results seem to imply that 
the above second candidate is the most promising for the progenitor
object of the M31's nuclear disk,
observational results have not been so accumulated as to
reveal clearly the nature of stellar population of M31's nucleus.
Thus more detailed spectroscopic studies that can clarify
the age and metallicity distribution
will provide valuable information on the origin of M31's nucleus.

\clearpage


\figcaption{
Mass distribution of a merger between
the central MBH and a stellar cluster   projected
onto $x$-$y$ plane (upper four panels) and onto $x-z$ one
(lower four panels) at each time $T$
for the fiducial model 
with $M_{\rm sat}/M_{\rm BH}$ = 0.1,  $M_{\rm gal}/M_{\rm BH}$
=0.3, $V_{\rm x}$ = $-0.5 V_{\rm cir}$, and $V_{\rm y}$ = 0.75$V_{\rm cir}$. 
$T$ (in our units)  is indicated in the upper
left-hand corner of each panel.
The MBH is initially located exactly in the center of 
a bulge component (i.e., the $x$ and $y$ position ($x$,$y$) is (0,0)
in this figure) and is not plotted in this figure for clarity. 
Although the MBH does  not move freely
from the initial position for $T$ $<$ 2.8,
it begins to slightly drift  after $T$ = 2.8 owing to dynamical
impact of the merging cluster.
The MBH's ($x$,$y$) is  (0.12,0.08) at $T$ = 3.2
and (0.09,0.14) at $T$ = 4.0. 
Here the  scale is given in our units (corresponding to 10 pc) and 
each of the  8  frames measures 25 pc (2.5 length units) 
on a side.   
Note that strong tidal field of the MBH transforms the
initial spherical shape of the cluster into a thick disk. 
\label{fig-1}}

\figcaption{
The upper left panel shows the  distribution of orbital
eccentricity ($e$) of stars in the developed nuclear thick disk.
Here f(e) represents the number fraction of stars for each eccentricity
bin at $T$ = 4.0.
The radial density profile, the rotation curve,  
and the velocity dispersion profile along $x$-axis (solid line)
and $y$-axis (dotted one) at $T$ = 4.0 for the disk
are given in the upper right
panel, in the lower left one, and in the lower right one,
respectively.
Here the scale for length and velocity is given in our units
and the surface density scale is given in units of 100 $\times$ 
$M_{\rm BH}$/${R_{\rm ini}}^2$ for clarity. 
Note that structure and kinematics of the disk clearly show
asymmetry in the radial distributions.
\label{fig-2}}

\figcaption{
Final mass distribution of the four models with
$M_{\rm sat}/M_{\rm BH}$ = 1.0 and $M_{\rm gal}/M_{\rm BH}$
= 3.0 (upper left, labeled as  A), with $V_{\rm x}$ = 0.0 and
$V_{\rm y}$ = $V_{\rm cir}$ (upper right, B),
with $V_{\rm x}$ = $-1.5$$V_{\rm cir}$ and
$V_{\rm y}$ = 0.75$V_{\rm cir}$ (lower left, C),
and with $V_{\rm x}$ = $-1.0$$V_{\rm cir}$
and $V_{\rm y}$ = 0.1$V_{\rm cir}$ (lower right, D),
all of which fail to form a nuclear thick disk around  the MBH.
Here parameter values other than those specified above
are exactly the same as those of the fiducial model
for each of the four models.
Fundamental characteristics of each model are summarized as follows.
Model A: Tidal gravitational field  of the MBH is rather weak
compared with the external gravitational field of the bulge.
B: Initially the orbit of the cluster is exactly circular 
so that the cluster can not become very close to the MBH.
C: The cluster encounters  with the MBH in a hyperbolic way
and does  not severely suffer from the MBH's tidal field
so that the cluster can escape from the MBH's with only the outer
part of the stellar components tidally stripped away from it
(and is well outside the frame at this time $T$ =4.0).
D: The cluster is initially in a very radial orbit (with small
pericenter distance).
\label{fig-3}}

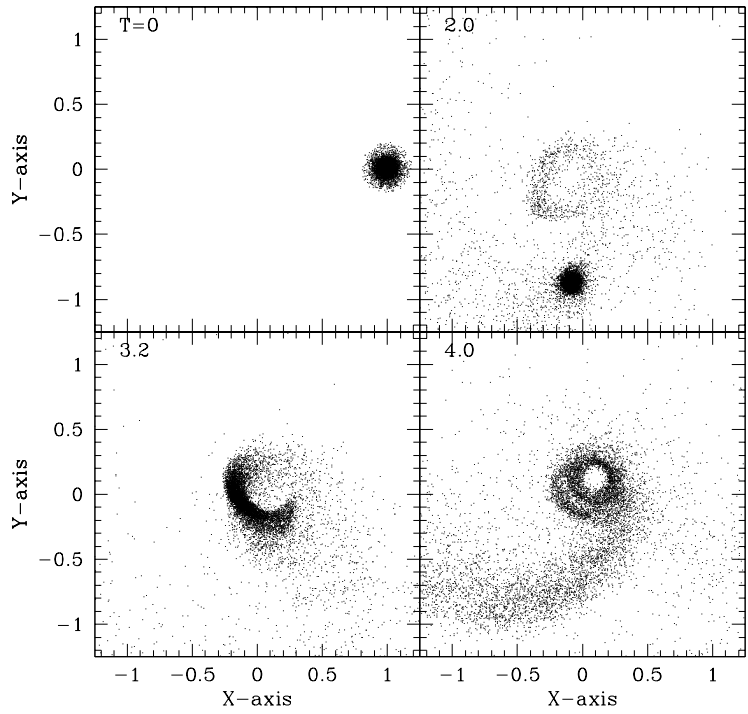
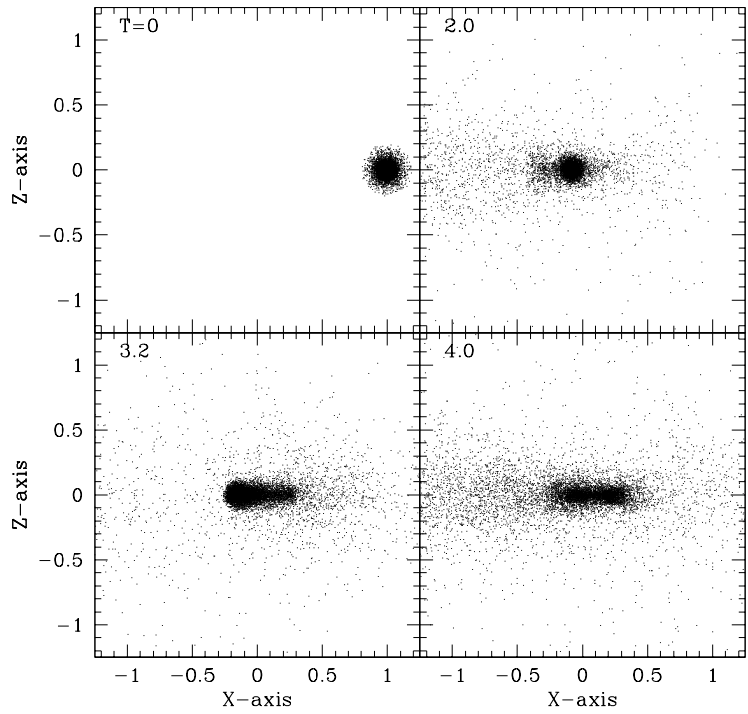
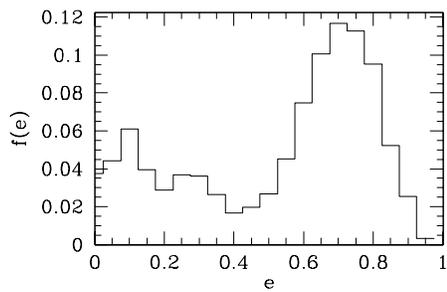
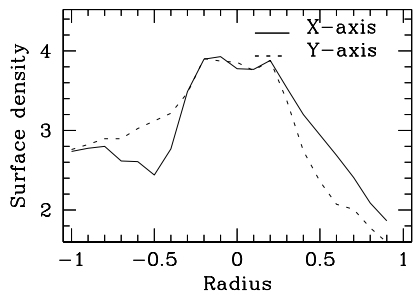
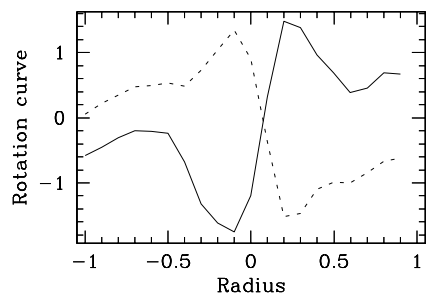
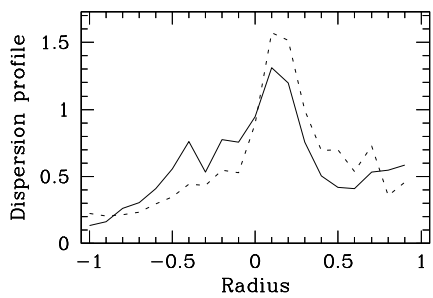
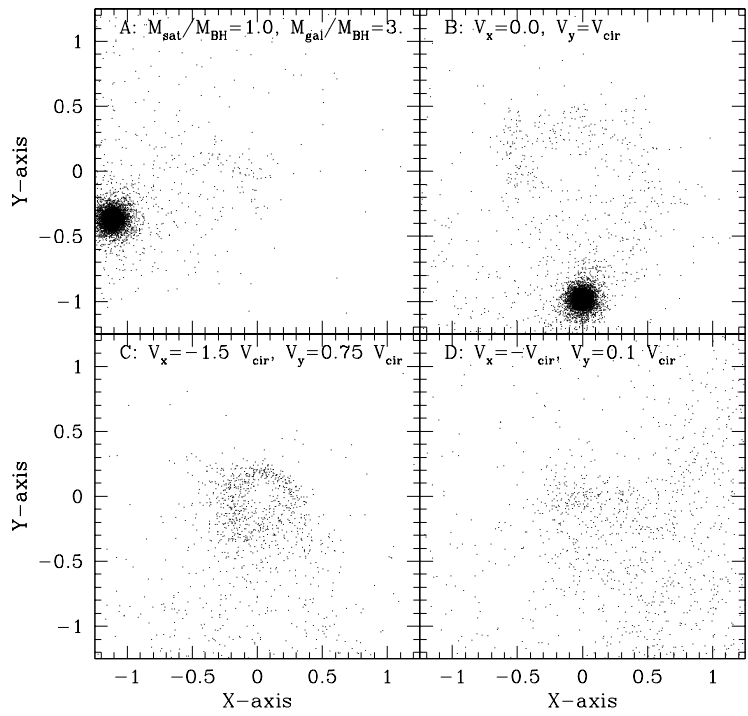
\end{document}